\documentclass[conference]{IEEEtran}
\IEEEoverridecommandlockouts
\usepackage{cite}
\usepackage{amsmath,amssymb,amsfonts}
\usepackage{graphicx}
\usepackage{textcomp}
\usepackage{xcolor}
\usepackage{tikz}
\usetikzlibrary{external}
\usepackage{gensymb}
\usepackage[nolist]{acronym}
\usepackage{algorithm}
\usepackage{forest}
\usepackage[noend]{algpseudocode}

\algnewcommand{\IIf}[1]{\State\algorithmicif\ #1\ \algorithmicthen}
\algnewcommand{\EndIIf}{\unskip\ \algorithmicend\ \algorithmicif}

\newcommand{\correction}[1]{\textcolor{black}{#1}}
\newcommand{\correctionwk}[1]{\textcolor{black}{#1}}

\def\BibTeX{{\rm B\kern-.05em{\sc i\kern-.025em b}\kern-.08em
    T\kern-.1667em\lower.7ex\hbox{E}\kern-.125emX}}
\begin{document}

	\begin{acronym}
	\acro{MCTS}{\textit{Monte Carlo Tree Search}}
	\acro{AoA}{\textit{Angle of Arrival}}
	\acro{SIR}{\textit{Sampling Importance Resampling}}
	\acro{ULA}{\textit{uniform linear array}}
	\acro{PDF}{\textit{probability density functions}}
	\acro{PMF}{\textit{probability mass function}}
	\acro{UCB}{\textit{Upper Confidence Bound}}
	\acro{ASN}{\textit{Acoustic Sensor Network}}
	\acro{SLAM}{\textit{Simultaneous Localization and Mapping}}
	\acro{alg.}{\text{Algorithm}}
\end{acronym}

\title{Active Acoustic Source Tracking Exploiting Particle Filtering and Monte Carlo Tree Search
}

\author{\IEEEauthorblockN{Thomas Haubner, Alexander Schmidt, and Walter Kellermann}
\IEEEauthorblockA{\textit{Multimedia Communications and Signal Processing,University of Erlangen-Nuremberg} \\
Cauerstr. 7, 91058 Erlangen, Germany,
thomas.haubner@fau.de}
}

\maketitle

\begin{abstract}
	In this paper, we address the task of active acoustic source tracking as part of robotic path planning. It denotes the planning of sequences of robotic movements to enhance tracking results of acoustic sources, e.g., talking humans, by fusing observations from multiple positions. Essentially, two strategies are possible: short-term planning, which results in greedy behavior, and long-term planning, which considers a sequence of possible future movements of the robot and the source. %
	Here, we focus on the second method as it might improve tracking performance compared to greedy behavior and propose a path planning algorithm which exploits \ac{MCTS} and particle filtering, based on a reward motivated by information-theoretic considerations. \correctionwk{By representing} \correction{the state posterior by weighted particles, we are capable of modelling arbitrary \ac{PDF}s and dealing with highly non-linear state-space models}.
\end{abstract}
\section{Introduction}
\label{sec:intro}
One of the main challenges in modern robotics is intuitive human-robot interaction \cite{intuitve_hum_rob_interaction} for which reliable information about the location of the human is essential. If microphones are mounted on the robot, e.g., for the purpose of \correctionwk{voice} communication, acoustic data can be used for inferring localization information. Typically, the relative \ac{AoA} between the robot and the source is estimated by, e.g., MUSIC \cite{music} or SRP-PHAT \cite{dibiase2000high}. Inferring distance information from microphone data is generally difficult in reverberant rooms and often relies on labeled training data, e.g., \cite{cdr_dist_est_brendel}.

\correctionwk{As an} option to improve localization accuracy \correctionwk{one can estimate} multiple \ac{AoA}s at different sensor positions in parallel and \correctionwk{fuse} them to obtain a position estimate. This can be achieved for example by exploiting several spatially distributed microphone arrays, i.e., an acoustic sensor network \cite{sensor_network}. %
Alternatively, \ac{AoA}s can also be measured sequentially, e.g., by a microphone array-equipped robot that is moving in a room \cite{act_mob_loc_vincent}. The latter approach greatly complicates the task as the robot-source configuration will generally vary over time. Thus, sequential fusing algorithms have to be employed.
%
On the bright side, however, sequential measurements \correctionwk{can exploit the controllability of the sensor topology of the robot relative to the target. Thereby, either the movement of the sensor array as a whole, or relative movements of the sensors within the array itself, e.g., by moving head or limbs of the robot, can be exploited such that target tracking can be supported.} This idea is referred to as active sensing.
%
%
%
%
For this problem, heuristic strategies have been suggested, e.g., moving a robotic head \cite{heuristic_head_moving} or its limbs \cite{lnt2014-38} furnished with microphones in the direction of the estimated source. Besides heuristic policies, different algorithms have been proposed to solve this task by maximizing a \correctionwk{specific} objective function\correction{, e.g., \cite{act_mob_loc_vincent}, \cite{long_term_mot_planning_mcts_vincent}, \cite{thomsen_heuristic_2015}} or \cite{kreuecher_mt_act_sensing_myopic}. \correctionwk{These} approaches can be categorized as greedy or long-term path planning algorithms. Greedy one-step algorithms, as proposed, e.g., in \cite{kreuecher_mt_act_sensing_myopic} and \cite{thrun_entr_act_loc}, model the objective to depend only on one future hypothesis. This short-term planning is prone to suboptimal tracking performance as it ignores measurements while moving the sensor and subsequent ones. Thus, long-term motion planning algorithms have been proposed in, e.g., \cite{act_mob_loc_vincent} for active localization, i.e., static sources, where the objective function is evaluated over several hypothetical future robot-source configurations.
Conditional independence of the objective function given prior movements is assumed to deal with the exponentially increasing amount of possible future hypotheses \cite{act_mob_loc_vincent}. 
In \cite{long_term_mot_planning_mcts_vincent}, an extension based on \ac{MCTS} has been proposed which does not rely on the previously assumed conditional independence. Moreover, to circumvent the limitation of discrete source positions \cite{act_mob_loc_vincent}, the state posterior is modelled by a mixture of Gaussians \cite{long_term_mot_planning_mcts_vincent}. 
\vspace*{-.1cm}

%
\correction{
In this paper, we introduce a path planning algorithm for active tracking which is also based on \ac{MCTS} but employs a sequential Monte Carlo method \cite{seq_mc_practice}, for updating the state posterior, i.e., belief. As sequential Monte Carlo methods approximate the ideal continuous Bayes filter \cite{prob_robotics}, we are capable of overcoming the limitations of Kalman filter based planning algorithms, e.g., the  approximation of the non-linearity by a Taylor series \cite{prob_robotics}, and address a broader class of non-linear movement models which are commonly \correctionwk{encountered} in acoustic source tracking applications \cite{evers_movement_model}. Representing the state posterior by weighted particles \correctionwk{also allows to model} arbitrary \ac{PDF}s, e.g., multi-modal \ac{PDF}s resulting from front-back ambiguity \cite{van2004optimum}. As planning objective we employ the expected information gain of applying a specific action, based on an estimate of the differential entropy of the belief \cite{pf_entropy}. Due to the very general description of the state posterior by weighted particles, we are also capable of exploiting higher-order statistics included in the planning objective. Furthermore, compared to \cite{act_mob_loc_vincent} and \cite{long_term_mot_planning_mcts_vincent}, uncertainty of robotic movements is taken into account during \correctionwk{planning. This} is decisive for real-world applications as the execution of the same control command in various scenarios leads to different movements depending on, e.g., the composition of the ground. Additionally, \correctionwk{while} \cite{act_mob_loc_vincent} and \cite{long_term_mot_planning_mcts_vincent} have simulated only static sources, we are focusing on moving sources \correctionwk{as a common scenario, e.g.,  when involving} talking humans. Finally, the efficacy of the proposed algorithm and the effect \correctionwk{of} different planning depths is shown by simulations.	
%
%
%
}

\section{Acoustic Source Tracking}
\label{sec:acoustic_source_tracking}
In the following section, the proposed system is described \correctionwk{by} a state-space model. Based on this, an appropriate tracking algorithm is discussed.

\subsection{State-Space Model}
\label{sec:state_space_mod}
%
The system state at time $t$, captured by the state vector $\boldsymbol{x}_t =\begin{pmatrix}(\boldsymbol{x}_t^{r})^\text{T} & (\boldsymbol{x}_t^s)^\text{T} \end{pmatrix}^\text{T}$, comprises the \correctionwk{robotic - i.e., sensor array - state} $\boldsymbol{x}_t^{r}$ and the source state $\boldsymbol{x}_t^{s}$.
For simplicity, the array and the source are assumed to move in the same 2D plane. A state describing either the array $q=r$ or the source $q=s$ is given by $\boldsymbol{x}^q_t = \begin{pmatrix}
x^q_t & y^q_t &  \theta^q_t & v^q_t
\end{pmatrix}^\text{T}$
%
including the position in Cartesian coordinates $\begin{pmatrix} x^q_t & y^q_t \end{pmatrix}^\text{T}$, orientation $\theta^q_t$ and speed $v^q_t$. At each time step the system can be influenced by a specific action $u_{q,t}$ \correction{controlling} the angular speed. Following \cite{evers_movement_model}, the system dynamics is modelled by the nonlinear \correction{constant-velocity movement model}
\begin{align}
\theta_t^q & = \text{wrap}_{[-180\degree,180\degree]}(\theta_{t-1}^q + u_{q,t} \Delta T + w_t^{q,\theta}), \label{eq:transition_equations_1}\\[.35em]
v_t^q &= v_{t-1}^q + w_t^{q,v},\label{eq:transition_equations_2}\\[.35em]
\begin{pmatrix}
x_t^q \\ y_t^q
\end{pmatrix} &= \begin{pmatrix}
x_{t-1}^q \\ y_{t-1}^q
\end{pmatrix} + \begin{pmatrix}
\cos(\theta_t^q) \\ \sin(\theta_t^q)
\end{pmatrix} \Delta T ~ v_t^q + \begin{pmatrix}
w_t^{q,x} \\ w_t^{q,y}
\end{pmatrix}.
\label{eq:transition_equations_3}
\end{align}
%
%
Hereby, $w_t^{(\cdot)} \sim \mathcal{N}(0, \sigma^2_{(\cdot)})$ denotes additive Gaussian noise\correction{, parametrized by the variances $\sigma_{q, x}^2$, $\sigma_{q, y}^2$, $\sigma_{q, v}^2$ and $\sigma_{q, \theta}^2$,} to account for model uncertainties. $\text{wrap}_{[a,b]}(\cdot)$ is a function which wraps its argument to the range $[a,b]$ and $\Delta T$ the sampling interval between successive observations. Note that only the robot can be controlled and thus $u_{s,t}=0\degree \frac{1}{\text{s}}$. The state equations \eqref{eq:transition_equations_1} - \eqref{eq:transition_equations_3} do not just account for uncertain source movements but also model the array movement to be noisy. Inference about the states is drawn from observations as follows: While we assume the array state to be known, its movement is stochastic. Information about the latent source states is obtained by observing the sound field as sensed by the robot's microphone array. Hence, under the assumption that the source is continuously active and located in the far-field, an estimate of the \ac{AoA} \mbox{${z}_t \in \mathcal{Z}$}, e.g., a finite uniform grid of \ac{AoA} hypotheses with a given angular resolution $\rho$, between the robot and the source can be computed from the microphone measurements at each time step. In this paper, the conditional measurement \ac{PMF} $p({z}_t | \boldsymbol{x}_t)$ of observing ${z}_t$ while being in state $\boldsymbol{x}_t$ is assumed to be a quantized Gaussian
%
%
\begin{align}
p(z_t | \boldsymbol{x}_t) = \int_{z_t - \frac{\rho}{2}}^{z_t+\frac{\rho}{2}}\mathcal{N}\Big(\check{z};&{\mu}_{\text{m}}\left(\boldsymbol{x}_t\right),\sigma^2_{\text{m}}\left(\boldsymbol{x}_t\right)\Big) d \check{z} \label{eq:obs_mod}
\end{align} 
with mean $\mu_{\text{m}}(\cdot)$ and variance $\sigma^2(\cdot)$ being typically non-linear functions of the state. These parameters have to be either modelled or learned. We suggest to learn them in a training phase. In \cite{act_mob_loc_vincent}, it was assumed that these parameters mainly depend on the distance $d(\boldsymbol{x})$ 
%
%
and the true \ac{AoA} $\phi_{\text{true}}(\boldsymbol{x})$
%
%
%
%
%
between the robot and the source.
However, if the robot is equipped with a \ac{ULA}, it can additionally be assumed that sources which are located symmetrically with respect to the array axis inherit similar parameters. 
%
%
%
\correction{These assumptions greatly simplify the task of learning the observation model as its mean and variance do not depend on the absolute states of the robot and the source anymore.}

\subsection{Bayesian Tracking}
\label{sec:bayesian_tracking}
As the states of the system are partially hidden, we introduce the state posterior, i.e., belief, $p(\boldsymbol{x}_t | \boldsymbol{h}_{t})$, representing the knowledge about the states, given the vector \mbox{$\boldsymbol{h}_t=\begin{pmatrix}u_{r,1}& z_1& \dots& u_{r,t}& z_{t}\end{pmatrix}^\text{T}$} of prior actions and observations \cite{prob_robotics}.
This leads to the question how to fuse a new action-observation vector $\begin{pmatrix}u_{r,t+1}& {z}_{t+1}\end{pmatrix}^{\text{T}}$ into a given belief. Straightforward extensions of the Kalman filter update, i.e., extended or unscented Kalman filter \cite{unscented_kf}, might not model accurately the belief due to the possibly multi-modal state posterior and highly nonlinear state transition equations. 
\correction{Thus, we employ a sequential Monte Carlo method \cite{seq_mc_practice}. Hereby, the belief is represented by a set \mbox{$\Gamma_{t} = \{(\boldsymbol{x}_t^{(1)}, \omega_t^{(1)}), \dots, (\boldsymbol{x}_t^{(I)}, \omega_t^{(I)}) \}$} of $I$ weighted particles whose elements are tuples of state samples $\boldsymbol{x}_t^{(i)}$ and weights $\omega_t^{(i)}$. 
%
%
In this paper a \ac{SIR} particle filter with systematic resampling is used to update the belief from $t$ to $t+1$ \cite{seq_mc_practice,gustafsson2010particle,resampling_douc} which is abstractly described by
\begin{equation}
\left(\Gamma_{t+1} \right) = \text{SIR}(\Gamma_{t}). \label{eq:pf_update}
\end{equation}
}
%
%
%

%
%
\section{Path Planning}
\label{sec:path_planning}
Paths are considered in this paper as consecutively executed control commands chosen from a finite set of $J$ discrete commands $u_{r,t} =u_{t} \in \mathcal{U}$. Thus, path planning can be interpreted as a sequential decision making problem, i.e., selecting the optimal action at each time step.
%
\vspace*{-.1cm}
\subsection{Rewards and Returns}
\label{sec:rewards_and_returns}
For comparing different paths and actions the concept of rewards and returns, as described in \cite{rl_sutton}, is adopted. Rewards $R(\boldsymbol{h}_t)$ can be interpreted as instantaneous feedback on the performance of a specific control command $u_t$. Ideally one wants to minimize an error norm between the true state and the estimated state. However, the true state is not known. Thus, a common idea is to minimize the belief uncertainty which has to be quantified. \cite{thrun_entr_act_loc} proposed the negative differential entropy of the belief as intrinsic measure of uncertainty for active localization. We adopt this idea by utilizing the recursive weighted particle-based entropy estimator proposed in \cite{pf_entropy} 
%
%
\begin{align}
R( &\boldsymbol{h}_t)=-\hat{\mathcal{H}}[ p(
\boldsymbol{x}_t| \boldsymbol{h}_{t})] = -\log \left( \sum_{i=1}^I p(z_t |\boldsymbol{x}_t^{(i)}) \omega^{(i)}_{t-1} \right) \label{eq:entropy_est_rew} \\
&+ \sum_{i=1}^{I} \log \left(p(z_t|\boldsymbol{x}_t^{(i)}) \left( \sum_{\nu=1}^{I} p(\boldsymbol{x}_t^{(i)}|\boldsymbol{x}_{t-1}^{(\nu)},u_t) \omega_{t-1}^{(\nu)} \right) \right) \omega_t^{(i)} \notag
\end{align}
%
%
\correction{corresponding to the history \mbox{$\boldsymbol{h}_t^{\text{T}} = \begin{pmatrix} \boldsymbol{h}_{t-1}^{\text{T}} & u_t & z_t \end{pmatrix}$}}.
%

As rewards can only account for a single action and observation at a time, the concept of returns 
%
\begin{equation}
G(\boldsymbol{h}_{t+K}) = \sum_{k=1}^K \gamma^{k-1} R(\boldsymbol{h}_{t+k}) \label{eq:ret_def}
\end{equation}
as cumulative discounted, i.e., weighted, \correction{future} rewards is introduced \cite{rl_sutton}. Hereby, $\gamma \in [0,1]$ defines the discount factor which \correctionwk{controls} the behavior of striving for early rewards within the planning horizon of $K$ time intervals.
%
%
%
%
As actions and observations are stochastic, the expected future return
\begin{equation}
	V(\boldsymbol{h}_t, u_{t+1}) = \mathbb{E}[G(\boldsymbol{h}_{t+K})|u_{t+1},\boldsymbol{h}_{t}],
\end{equation}
starting from history $\boldsymbol{h}_t$ and selecting action $u_{t+1}$, is introduced \cite{rl_sutton}. The action maximizing the expected future return is selected for path planning 
%
%
\begin{equation}
\vspace*{-.2cm}
u^{\text{opt}}_{t+1} = \underset{u \hphantom{.} \in \hphantom{.} \mathcal{U}}{\text{argmax}} ~V(\boldsymbol{h}_t, u).
\end{equation}
%
%
%

\vspace{-.025cm}
\subsection{Monte Carlo Tree Search Planning}
\label{sec:MCTS}
%
%
%
%
\begin{figure}[tb]
	\centering
	\tikzset{action1/.style={circle,  draw, inner sep=3pt,draw}}
	\tikzset{action2/.style={circle, fill=black,inner sep=3pt,draw}}
	\begin{forest}
		for tree={
			s sep=30pt,
			l sep= 15pt
		},
		my u/.style n args=3{
			edge label={node [midway, #3, text=black] {$~u_{#1}^{#2}~$}},
		},
		my z/.style n args=3{
			edge label={node [midway, #3, text=black] {$~z_{#1}^{#2}~$}},
		}
		[$\boldsymbol{h}_t$
		[,action1, my u={t+1}{1}{above left}
		[,action2, my z={t+1}{1}{left}
		[,action1, my u={t+2}{1}{left}
		[,action2, my z={t+2}{1}{left}]
		]
		]
		[,action2, my z={t+1}{2}{below left}
		[,action1, my u={t+2}{2}{left}
		[,action2, my z={t+2}{1}{left}
		]
		[,action2 , my z={t+2}{2}{right}]
		]
		]
		[,action2, my z={t+1}{3}{right}]
		]
		[,action1, my u={t+1}{2}{above right}
		[,action2, my z={t+1}{1}{left}]	
		[,action2, my z={t+1}{3}{right}]
		]
		]
	\end{forest}
	\caption{\correction{Partial tree $\mathcal{T}_{\boldsymbol{h}_t}$ with action set $\mathcal{U}= \{u^{1},u^{2}  \}$ and observation set $\mathcal{Z}=\{ z^{1},z^{2},z^{3}\}$}}
	\vspace*{-.05cm}
	\label{fig:mc_hist_tree}
\end{figure}
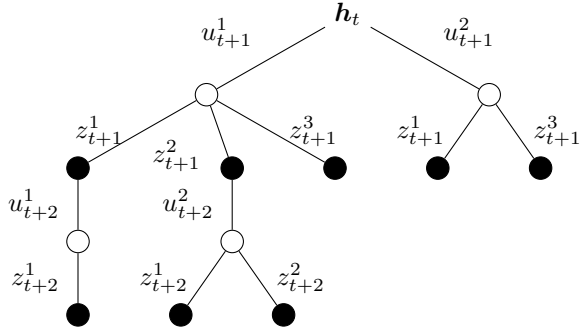
%
\correction{
\correctionwk{Now we address the problem of} how to efficiently compute the expected future returns when applying the actions $u_{t+1} \in \mathcal{U}$: \correctionwk{One} option is to use Monte Carlo simulation, i.e., simulating $L$ returns corresponding to hypothetical sequences of future actions and observations $\boldsymbol{h}_{t+K} = \begin{pmatrix}  \boldsymbol{h}_t^{\text{T}} ~ \dots~ u_{t+K}~ {z}_{t+K}\end{pmatrix}^{\text{T}}$ followed by arithmetic averaging \correctionwk{over all trials}. However, this approach \correctionwk{suffers from the fact that} the number of possible future hypotheses is growing exponentially with increasing planning horizon $K$. To overcome this problem\correctionwk{,} the \ac{MCTS} technique  \cite{mcts_survey} has been developed which builds up a partial tree $\mathcal{T}_{\boldsymbol{h}_t}$\correction{, i.e., a set of simulated future hypotheses}, as exemplified in Figure \ref{fig:mc_hist_tree} for the action set $\mathcal{U}= \{u^{1},u^{2} \}$ and observation set $\mathcal{Z}=\{ z^{1},z^{2},z^{3}\}$, and \correction{selects appropriate actions} to concentrate on promising paths, i.e., individual sequences with high returns. The exploitation-exploration dilemma, encountered during the sequential decision-making \correctionwk{on} which paths to investigate, is hereby tackled by the \ac{UCB} policy \cite{kocsis2006improved} \correctionwk{formulated as}
}
\begin{equation}
u^{\text{UCB}}_{t+1} = \underset{u\hspace*{.05cm} \in ~\mathcal{U}}{\text{argmax}}~ \left(\hat{V}(\boldsymbol{h}_{t}, u) + c \sqrt{\frac{\ln N(\boldsymbol{h}_t)}{N_u(\boldsymbol{h}_t, u) + \epsilon}} \right). \label{eq:ucb_pol}
\end{equation}
\correction{
Hereby, $N(\boldsymbol{h}_t)$ denotes the number of previous visits of the simulated hypothesis $\boldsymbol{h}_t$, $N_u(\boldsymbol{h}_t, u)$ \correctionwk{is} the number of previously simulated situations where action $u\in\mathcal{U}$ had been selected \correction{and $\epsilon$ \correctionwk{is} a small number to prevent division by $0$}. The ratio of exploration and exploitation can be controlled via the hyperparameter \correction{$c>0$} \cite{mcts_survey}.
}
\begin{algorithm}[tb]
	\caption{Monte Carlo path planning at time step $t$}
	\label{alg:mc_path_plan}
	\begin{algorithmic}[1]
		\For{$i=1:L$}
		\State Sample particle $\boldsymbol{x}_t^{(i)}$ from density defined by $\Gamma_{t}$
		\State \Call{Sim}{$\boldsymbol{x}_t^{(i)}, \boldsymbol{h}_{t},\mathcal{T}_{\boldsymbol{h}_t}, 0$}
		\EndFor
		\State $u^{\text{opt}}_{t+1} = \underset{u\hspace*{.05cm}  \in ~\mathcal{U}}{\text{argmax}}~ \hat{V}\left(\boldsymbol{h}_{t}, u\right)$ 
				\State Move robot by executing $u_{t+1}^{\text{opt}}$
				\State Measure \ac{AoA} $\tilde{z}_{t+1}$	
				\State $\boldsymbol{h}_{t+1}^{\text{T}} = \begin{pmatrix} \boldsymbol{h}_t^{\text{T}} &u_{t+1}^{\text{opt}} & \tilde{z}_{t+1} \end{pmatrix}$
				\If {$\boldsymbol{h}_{t+1} \notin \mathcal{T}_{\boldsymbol{h}_t}$} $\Gamma_{{t+1}} = \text{SIR}(\Gamma_{t})~$ \hspace*{.8cm}(see \eqref{eq:pf_update})
				\EndIf
				\State $\mathcal{T}_{\boldsymbol{h}_{t+1}} \gets\text{prune }(\mathcal{T}_{\boldsymbol{h}_{t+1}})$
				\State $t \gets t+1$
	\end{algorithmic}
\end{algorithm}

\correction{Algorithm \ref{alg:mc_path_plan}, \ref{alg:mc_sampling_procedure} and \ref{alg:def_pol_alg} jointly describe the proposed path planning method, using \ac{MCTS} and particle filtering. \correctionwk{The main Algorithm~\ref{alg:mc_path_plan} samples $L$ times from the current belief (\textit{line}~(l.)~2,~Alg.~\ref{alg:mc_path_plan}) and invokes Algorithm \ref{alg:mc_sampling_procedure}, i.e., function \textproc{sim}, for updating the estimated future returns $\hat{V}(\boldsymbol{h}_{t}, \cdot)$ by simulating possible future} hypotheses $\boldsymbol{h}_{t+K}$ (l.~3,~Alg.~\ref{alg:mc_path_plan}), \correctionwk{including the computation of} the corresponding returns $G(\boldsymbol{h}_{t+K})$ and arithmetic averaging, i.e., (l.~15,~Alg.~\ref{alg:mc_sampling_procedure}). These simulated future hypotheses are depicted in Figure \ref{fig:mc_hist_tree} as various paths starting from history $\boldsymbol{h}_t$. As only the first new action-observation sequence is added to the tree $\mathcal{T}_{\boldsymbol{h}_t}$ (l.~11~Alg.~\ref{alg:mc_sampling_procedure}), the respective hypotheses have varying lengths. After planning, the robot moves by executing the optimal control command $u_{t+1}^{\text{opt}}$ (l.~5~Alg.~\ref{alg:mc_path_plan}) and measures the real \ac{AoA} $\tilde{z}_{t+1} \in \mathcal{Z}$ (l.~6~Alg.~\ref{alg:mc_path_plan}). If the true hypothesis is not part of the tree, the state posterior is updated (l.~8~Alg.~\ref{alg:mc_path_plan}). Subsequently, the tree is pruned (l.~9~Alg.~\ref{alg:mc_path_plan}), i.e., all hypotheses which do not include $\boldsymbol{h}_{t+1}$ are deleted, and the planning starts again.
}
\begin{algorithm}[tb]
	\caption{Monte Carlo Return Estimation \correctionwk{- \textproc{sim}}}
	\label{alg:mc_sampling_procedure}
	\begin{algorithmic}[1]
		\Function{sim}{$\boldsymbol{x}_k^{(i)}, \boldsymbol{h}_k, \mathcal{T}_{\boldsymbol{h}_t}, k$}
		\If {$k\geq K$} {\bfseries return} $0$ \EndIf
		\State $u_{k+1}^{\text{UCB}} = \underset{u \hspace*{.05cm} \in ~\mathcal{U}}{\text{argmax}}~ \left(\hat{V}(\boldsymbol{h}_k, u) + c \sqrt{\frac{\ln N(\boldsymbol{h}_k)}{N_u(\boldsymbol{h}_k,u)+\epsilon}} \right)$ 
		\State $\boldsymbol{x}_{k+1}^{(i)} \sim p (\boldsymbol{x} | \boldsymbol{x}_{k}^{(i)}, u_{k+1}^{\text{UCB}} )~$ (see \eqref{eq:transition_equations_1}, \eqref{eq:transition_equations_2} and \eqref{eq:transition_equations_3})
		\State ${{z}}_{k+1} \sim p({z}|\boldsymbol{x}_{k+1}^{(i)})~$ \hspace*{.8cm}(see \eqref{eq:obs_mod})
		\State $\boldsymbol{h}_{k+1}^{\text{T}} = \begin{pmatrix} \boldsymbol{h}_{k}^{\text{T}} &u_{k+1}^{\text{UCB}} & z_{k+1} \end{pmatrix}$
		\If {$\boldsymbol{h}_{k+1} \in \mathcal{T}_{\boldsymbol{h}_t}$}
		\State $\tilde{G} =R(\boldsymbol{h}_{k+1}) + \gamma  $\Call{sim}{$\boldsymbol{x}_{k+1}^{(i)}, \boldsymbol{h}_{k+1}, \mathcal{T}_{\boldsymbol{h}_{t}}, k+1}$
		\Else
		\State $\Gamma_{{t+1}} = \text{SIR}(\Gamma_{t})~$ \hspace*{.8cm}(see \eqref{eq:pf_update})
		\State $\mathcal{T}_{\boldsymbol{h}_t} \gets \mathcal{T}_{\boldsymbol{h}_{t}} \cup \left\{ \boldsymbol{h}_{k+1} \right\}$
		\State $\tilde{G} =R(\boldsymbol{h}_{k+1}) + \gamma  $\Call{def}{$\boldsymbol{x}_{k+1}^{(i)}, \boldsymbol{h}_{k+1},k+1$}
		\EndIf
		\State $N(\boldsymbol{h}_{k}) \gets N(\boldsymbol{h}_{k})+1$ 
		\State $N_u(\boldsymbol{h}_{k},u_{k+1}^{\text{UCB}}) \gets N_u(\boldsymbol{h}_{k},u_{k+1}^{\text{UCB}})+1$
		\State $\hat{V}(\boldsymbol{h}_{k}, u_{k+1}^{\text{UCB}}) \gets (1-\frac{1}{N_u(\boldsymbol{h}_k,u_{k+1}^{\text{UCB}})})\hat{V}(\boldsymbol{h}_{k}, u_{k+1}^{\text{UCB}})$ \hspace*{3.5cm}$+ \frac{\tilde{G}}{N_u(\boldsymbol{h}_k,u_{k+1}^{\text{UCB}})} $
		\State {\bfseries return} $\tilde{G}$
		\EndFunction
	\end{algorithmic}
\end{algorithm}
\correction{
\textproc{sim} (Alg.~\ref{alg:mc_sampling_procedure}) is a recursive algorithm  which builds up the partial \correctionwk{tree} $\mathcal{T}_{\boldsymbol{h}_t}$ consisting of the respective future movements and observations and computes the estimated expected future returns. \correctionwk{If} the maximum planning depth $K$ is \correctionwk{not yet} reached, a future hypothesis is simulated \correctionwk{in Alg.~\ref{alg:mc_sampling_procedure}} by initially selecting an action $u^{\text{UCB}}_{k+1}$ according to the \ac{UCB} policy 
(l.~3~Alg.~\ref{alg:mc_sampling_procedure}). \correction{The selection of actions is visualized in Figure \ref{fig:mc_hist_tree} by either choosing $u_{t+k}^1$ or $u_{t+k}^2$.} Subsequently, a future state $\boldsymbol{x}^{(i)}_{k+1}$ is sampled (l.~4~Alg.~\ref{alg:mc_sampling_procedure}) from the \ac{PDF} described by the state transition equations \mbox{\eqref{eq:transition_equations_1} - \eqref{eq:transition_equations_3}} conditioned on the sampled particle and selected action. Then, an observation ${z}_{k+1}$ is sampled from the measurement process (l.~5~Alg.~\ref{alg:mc_sampling_procedure}). \correction{Sampling from the measurement process manifests itself in Figure \ref{fig:mc_hist_tree} by branching after the simulated actions, i.e., from the empty to the filled circles.} Based on the selected action $u^{\text{UCB}}_{k+1}$ and sampled measurement ${z}_{k+1}$, it is checked \correctionwk{whether} the simulated future hypothesis $\boldsymbol{h}_{k+1}$ is part of the tree $\mathcal{T}_{\boldsymbol{h}_t}$. If it is, the algorithm computes the return, i.e., recursive computation of \eqref{eq:ret_def}, by first computing the reward \eqref{eq:entropy_est_rew} and afterwards recursively calling itself with the new particle $\boldsymbol{x}_{k+1}^{(i)}$ and incremented depth $k$ (l.~8~Alg.~\ref{alg:mc_sampling_procedure}). If $\boldsymbol{h}_{k+1}$ has not been simulated \correctionwk{before}, the state posterior is updated (l.~10~Alg.~\ref{alg:mc_sampling_procedure}) and $\boldsymbol{h}_{k+1}$ is added to the tree $\mathcal{T}_{\boldsymbol{h}_t}$ (l.~11~Alg.~\ref{alg:mc_sampling_procedure}). \correctionwk{Subsequently, the following unexplored part of the tree is investigated by a default algorithm \textproc{def} which is applied until reaching the planning horizon $K$ (l.~12~Alg.~\ref{alg:mc_sampling_procedure}) \cite{mcts_survey}.} After simulating one future return $\tilde{G}$, $N(\boldsymbol{h}_t)$ and $N_{\text{UCB}}(\boldsymbol{h}_t)$ are incremented (l.~13~Alg.~\ref{alg:mc_sampling_procedure}, l.~14~Alg.~\ref{alg:mc_sampling_procedure}) and the expected future return $\hat{V}(\boldsymbol{h}_k, u_{k+1}^{\text{UCB}})$ is updated by computing the mean of the previously simulated returns and the current return $\tilde{G}$ (l.~15~Alg.~\ref{alg:mc_sampling_procedure}).
\begin{algorithm}[b]
	\caption{Default Policy \correctionwk{- \textproc{def}}}
	\label{alg:def_pol_alg}
	\begin{algorithmic}[1]
		\Function{def}{$\boldsymbol{x}_k^{(i)}, \boldsymbol{h}_k,  k$}
		\If {$k\geq K$} {\bfseries return} $0$ \EndIf
		\State Select $u_{k+1}^{\text{DEF}}$ 
		\State $\boldsymbol{x}_{k+1}^{(i)} \sim p (\boldsymbol{x} | \boldsymbol{x}_{k}^{(i)}, u_{k+1}^{\text{DEF}} )~$ (see \eqref{eq:transition_equations_1}, \eqref{eq:transition_equations_2} and \eqref{eq:transition_equations_3})
		\State ${{z}}_{k+1} \sim p({z}|\boldsymbol{x}_{k+1}^{(i)})~$ \hspace*{.8cm}(see \eqref{eq:obs_mod})
		\State $\boldsymbol{h}_{k+1}^{\text{T}} = \begin{pmatrix} \boldsymbol{h}_{k}^{\text{T}} &u_{k+1}^{\text{DEF}} & z_{k+1} \end{pmatrix}$
		\State {\bfseries return} $R(\boldsymbol{h}_{k+1}) + \gamma  $\Call{Def}{$\boldsymbol{x}_{k+1}^{(i)}, \boldsymbol{h}_{k+1},k+1$}
		\EndFunction
	\end{algorithmic}
\end{algorithm}
\newline
The default Algorithm~\ref{alg:def_pol_alg} \correctionwk{, i.e., function \textproc{def},} consists of computing an action $u_{k+1}^{\text{DEF}}$ according to a default policy (l.~3~Alg.~\ref{alg:def_pol_alg}), followed by sampling the transition (l.~4~Alg.~\ref{alg:def_pol_alg})  and observation \ac{PDF} to create possible future hypotheses (l.~5~Alg.~\ref{alg:def_pol_alg}) and recursively calling itself (l.~7~Alg.~\ref{alg:def_pol_alg}) with incremented depth. However, compared to \textproc{sim} the simulated hypothesis is not added to the tree $\mathcal{T}_{\boldsymbol{h}_t}$. The default policy cannot rely on experience from previous simulations for action selection \cite{mcts_survey}. Thus, in \cite{long_term_mot_planning_mcts_vincent} uniform random \correction{sampling from the action set $\mathcal{U}$ has been} used. We propose to incorporate the prior knowledge that moving closer to the target enhances localization results. However, in order to not fully rely on this naive assumption, our proposed default policy randomly uses either a uniform random behavior or an informed action with equal probability. The informed \correctionwk{decision is made by sampling from the \ac{PDF} defined by} the array transition equations \eqref{eq:transition_equations_1} - \eqref{eq:transition_equations_3} with $q=r$ for each possible action and selecting the one moving closest to the estimated source position. Subsequently, the default algorithm recursively calls itself until reaching the planning horizon $K$ and gives back the simulated return $\tilde{G}$. 
}
\section{Results}
\label{sec:results}
\correction{In this section, we evaluate the proposed planning algorithm in a simulated environment. Sensor and source are modelled to move at a constant speed of $v_t^r =v_t^s = 0.3~\frac{\text{m}}{\text{s}}$. The action set $\mathcal{U}$ of the sensor consists of $4$ different actions, namely staying in the current state, or changing the angular velocity $u_t$ to $ -45~\degree \frac{1}{\text{s}}$, $0~\degree \frac{1}{\text{s}}$ or $45~\degree \frac{1}{\text{s}}$. The respective standard deviations of the sensor and source movement models, i.e., \eqref{eq:transition_equations_1} - \eqref{eq:transition_equations_3}, are summarized in \mbox{Table \ref{tab:parameters_table}}. \correctionwk{The} control of the angular sensor movements is modelled to be noisy and thus taken into account for path planning. In contrast, the source movement model, i.e., $q=s$, \correctionwk{allows for} changes of velocity and orientation. The sampling interval $\Delta T$ is set to $1$ s. The particle filter state-space model was chosen according to the simulated model with additional source position uncertainties $\sigma_{s, x}= \sigma_{s, y}=0.1 \text{ m}$. The number of particles was set to $I=1000$.}
\begin{table}[tbp]
	\centering
	\caption{\correction{Parameters of the state-space model}}
	\label{tab:parameters_table}
	\begin{tabular}{l c c|c c c}
		& &$\sigma_{q, x}=\sigma_{q, y} ~[\text{m}]$ & $\sigma_{q,v}~[\frac{\text{m}}{\text{s}}]$ & $\sigma_{q, \theta}~[\degree\frac{1}{\text{s}}]$ \\
		\hline
		Sensor & ($q=r$) & $0$& $0$ & $5$ \\
		Source & ($q=s$) & $0$ & $0.025$& $10$ 
	\end{tabular}
\end{table}

The sensor consists of a \ac{ULA} with $6$ microphones and a spacing of $4.2$ cm. \ac{AoA}s are estimated by SRP-PHAT \cite{dibiase2000high} from microphone signals sampled at a sampling frequency of $8$~kHz and observation length of $4096$ samples per measurement. 
The observation parameters $\mu_{\text{m}}(\cdot, \cdot)$ and $\sigma^2_{\text{m}}(\cdot, \cdot)$ are estimated in a training phase by simulating various \ac{AoA} measurements, i.e., different sensor and source states, for each position of a uniform distance-\ac{AoA} grid \correction{with spacing $0.5$ m and $5 \degree$}. Positions in between are \correction{linearly interpolated}. Several rooms with dimensions $7 \text{m}\times 5 \text{m}\times 3\text{m} $, and varying reverberation time of $0.4$ s, $0.6$ s and $0.8$ s and SNR $10$ dB, $20$ dB and $30$ dB, are simulated \cite{allen1979image, habets2010room}. 
\begin{figure}[tb]
	\vspace*{-.5cm}
	\centering
	\resizebox{\columnwidth}{!}{\input{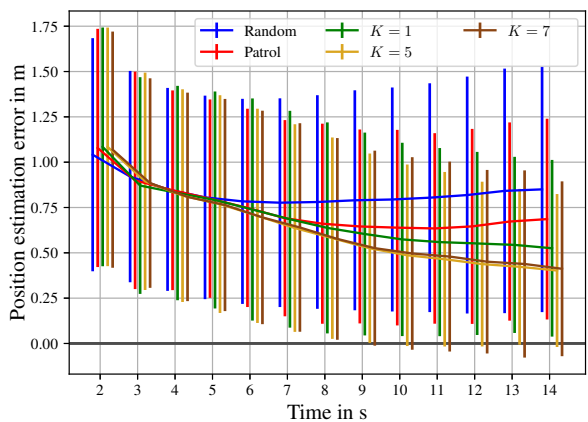}}
	\caption{Mean and standard deviation of estimation errors for various planning horizons \correction{$K$} \correctionwk{in comparison to random sensor movement and patrolling}}
	\label{fig:state_est_error_res}
	\vspace*{-.1cm}
\end{figure}
The path planning parameters are set to $\gamma=0.9$\correction{, cf.~\eqref{eq:ret_def},} and $c=1.75$\correction{, cf. \eqref{eq:ucb_pol}} \correction{to strive for long-term planning and to rely on previously simulated returns}. \correction{As suggested in \cite{mcts_survey}, the} rewards are normalized to lie in the interval $[-1,0]$. The maximum number of simulated \correctionwk{hypotheses} $L$ is set to $500$.
For reliable results, $300$ source paths and robot starting positions are created with uniformly distributed random initial positions. The initial source belief is modelled by particles whose positions and orientations are distributed uniformly in the enclosure. 
The first two actions are set to $u_{0}= u_{1} = 0\degree \frac{1}{\text{s}}$ to start path planning from an informed belief. 
The Euclidean distance between the true and the estimated source position\correction{, obtained by the weighted mean of the respective particles,} is computed for each time step and averaged over all $2700$ combinations of paths and rooms, i.e., reverberation times and SNRs. \correction{Figure~\ref{fig:state_est_error_res} depicts mean and standard deviation of this error over time, starting from $t=2$, i.e., after the two fixed actions at the beginning, for varying planning depths, i.e., $K=1$, $5$ and $7$. Additionally, as reference movement behaviour we simulated random action selection, similar to \cite{act_mob_loc_vincent,long_term_mot_planning_mcts_vincent}, and patrolling\correctionwk{, e.g., \cite{4209130},} the room by always moving to the most distant unexplored corner.}

\correction{ \correctionwk{From Figure~\ref{fig:state_est_error_res}  it becomes obvious that path planning, }either by patrolling or the proposed algorithm, significantly increases the accuracy and reliability of position estimates, compared to selecting an action at random at each time step. Additionally, greedy one-step planning ($K=1$) outperforms the the patrolling reference strategy. The effect of planning depths can be seen by comparing $K=1$ and $K=5$. Both perform similar at the beginning, however drift apart in the long-term. This is consistent with the idea that some actions might pay off later. However, a saturation effect can be seen. Planning $K=7$ time steps ahead does not improve the results further which can be explained either by the increased uncertainty of future predictions or an insufficient number of simulated paths. As the same effect has also been observed for $L = 1500$, we suppose this results from the increased uncertainty.}

%

\section{Summary}
\label{sec:summary_results}
\correction{In this paper, we proposed a path planning algorithm exploiting \ac{MCTS} \cite{long_term_mot_planning_mcts_vincent} and particle filtering. Based on the belief representation by weighted particles, an estimate of the expected differential entropy has been used as planning objective.} Additionally, a new default policy for \ac{MCTS} has been introduced. The efficacy of the algorithm and the effect of different planning horizons were demonstrated by simulated data of dynamic speech sources for a variety of challenging acoustic scenarios. 

\bibliographystyle{IEEEbib}
\bibliography{references}

\end{document}